\documentclass[journal]{IEEEtran}
\usepackage{color,amsfonts,amsmath,amssymb,booktabs,multirow,psfrag,bm,amsbsy,cite}
\usepackage{enumerate}
\usepackage{subfigure}
\usepackage[pdftex]{graphicx}
\usepackage{epstopdf}
  \DeclareGraphicsExtensions{.pdf,.jpeg,.png}

\begin{document}

\title{A Measurement Based Shadow Fading Model for Vehicle-to-Vehicle Network Simulations\thanks{This work was partially funded by the ELLIIT- Excellence Center at Link\"{o}ping-Lund In Information Technology and partially funded by Higher Education Commission (HEC) of Pakistan.}\thanks{T. Abbas, F. Tufvesson, and J. Karedal are with the Department of Electrical and Information Technology, Lund University, Lund, Sweden (e-mail: taimoor.abbas@gmail.com; Fredrik.Tufvesson@eit.lth.se; Johan.Karedal@ericsson.com).}\thanks{K. Sj\"{o}berg was a PhD student at the Centre for Research on Embedded Systems, Halmstad University, Halmstad, and now she is with the Department of Advanced Technology and Research at Volvo Group Trucks Technology (GTT), G\"{o}teborg, Sweden (e-mail: katrin.sjoberg@volvo.com).}}
\author{Taimoor Abbas, Katrin Sj\"{o}berg, \\ Johan Karedal and Fredrik Tufvesson}


\maketitle
\begin{abstract}
The vehicle-to-vehicle (V2V) propagation channel has significant implications on the design and performance of novel communication protocols for vehicular \emph{ad hoc} networks (VANETs). Extensive research efforts have been made to develop V2V channel models to be implemented in advanced VANET system simulators for performance evaluation. The impact of shadowing caused by other vehicles has, however, largely been neglected in most of the models, as well as in the system simulations. In this paper we present a shadow fading model targeting system simulations based on real measurements performed in urban and highway scenarios. The measurement data is separated into three categories, line-of-sight (LOS), obstructed line-of-sight (OLOS) by vehicles, and non line-of-sight due to buildings, with the help of video information recorded during the measurements. It is observed that vehicles obstructing the LOS induce an additional average attenuation of about $10$\,dB in the received signal power. An approach to incorporate the LOS/OLOS model into existing VANET simulators is also provided. Finally, system level VANET simulation results are presented, showing the difference between the LOS/OLOS model and a channel model based on Nakagami-$m$ fading.
\end{abstract}


\section{Introduction}
\label{sec:Introduction}

Vehicle-to-Vehicle (V2V) communication allows vehicles to communicate directly with minimal latency. The primary objective with the message exchange is to improve active on-road safety and situation awareness, e.g., collision avoidance, traffic re-routing, navigation, etc. The propagation channel in V2V networks is significantly different from that in cellular networks because V2V employs an \emph{ad hoc} network topology, both transmitter (TX) and receiver (RX) are highly mobile, and TX/RX antennas are situated on approximately the same height and close to the ground level. Thus, to develop an efficient and reliable system a deep understanding of V2V channel characteristics is required \cite{Javier2010}. 

A number of V2V measurements have been performed to study the statistical properties of V2V propagation channels \cite{Acosta07,Matolak08,Alex-09,Otto09,Oliver10}. Signal propagation over the wireless channel is often divided by three statistically independent phenomena named deterministic path loss, small-scale fading, and large-scale or shadow fading \cite{molisch05}. Path loss is the expected (mean) loss at a certain distance compared to the received power at a reference distance. The signal from the TX can reach the RX via several propagation paths or the multi-path components (MPC), which have different amplitudes and phases. The change in the signal amplitude due to constructive or destructive interference of the different MPCs is classified as small-scale fading. Finally, obstacles in the propagation paths of one or more MPCs can cause large attenuation and the effect is called shadowing. Shadowing gives rise to large-scale fading and it occurs not only for the line-of-sight (LOS) component but also for any other major MPC. Understanding all of these phenomena is equally important to characterize the V2V propagation channel. 

In real scenarios there can be light to heavy road traffic, involving vehicles with variable speeds and heights, and there are sometimes buildings around the roadside. Hence, it might be the case that the LOS is partly or completely blocked by another vehicle or a house. The received power depends very much on the propagation environment, and the availability of LOS. Moreover, in \cite{Taimoor11} it is reported that, in the absence of LOS, most of the power is received by single bounce reflections from physical objects. Therefore, for a realistic simulation and performance evaluation it is important that the channel parameters are separately characterized for LOS and non-LOS conditions. 

A number of different V2V measurement based studies with their extracted channel parameters are summarized in \cite{molisch09-CommMag}. For most of the investigations mentioned in \cite{molisch09-CommMag}, it is assumed that the LOS is available for the majority of the recorded snapshots. Thus, the samples from both the LOS and non-LOS cases are lumped together for modeling, which is somewhat unrealistic, especially for larger distances. The LOS path being blocked by buildings greatly impacts the reception quality in situations when vehicles are approaching the street intersection or road crossings. The buildings at the corners influence the received signal not only by blocking the LOS but also they act as scattering points which helps to capture more power in the absence of LOS \cite{TaimoorVTM2013}. A few measurement results for a non-LOS (NLOS) environment are available \cite{Mangel011,Giordano10,Mangel011-3,PKMWKP11,Turkka11} in which the path loss model is presented for different types of street crossings. 

In addition to the NLOS situation, the impact of neighboring vehicles can not be ignored. In \cite{Otto09}, it is reported that the received signal strength degrades on the same patch of an open road in heavy traffic hours as compared to when there is light traffic. These observed differences can only be related to other vehicles obstructing LOS since the system parameters remained the same during the measurement campaign. Similarly, Zhang \emph{et~al.} in \cite{Zang:2005} presented an abstract error model in which the LOS and NLOS cases are separated using a thresholding distance. It is stated that the signals will experience more serious fading in crowded traffic scenario when the distance between the TX and RX is larger than the thresholding distance.
 
In \cite{Boban11} and \cite{Meireles10}, it is shown that the vehicles (as obstacles) have a significant impact on LOS obstruction in both dense and sparse vehicular networks, implying that shadowing caused by other vehicles cannot be ignored in V2V channel models. To date, in majority of the findings for V2V communications except \cite{Boban11} and \cite{Meireles10}, the shadowing impact of vehicles has largely been neglected when modeling the path loss. It is important to model vehicles as obstacles, ignoring this can lead to an unrealistic assumptions about the performance of the physical layer, both in terms of received signal power as well as interference levels, which in turn can effect the behavior of higher layers of V2V systems. 

In order to characterize the channel parameters separately for LOS and non-LOS conditions V2V communication links in this paper are categorized into following three groups:
\begin{itemize}
\item Line-of-sight (LOS) is the situation when there is an optical line-of-sight between the TX and the RX. 
\item Obstructed-LOS (OLOS) is the situation when the LOS between the TX and RX is obstructed completely or partially by another vehicle.
\item Non-LOS (NLOS) is the situation when a building between the TX and RX completely block the LOS as well as many other significant MPCs. 
\end{itemize}

The channel properties for LOS, OLOS and NLOS are distinct, and their individual analysis is required. No path loss model, except a geometry based channel model published recently \cite{Boban2013arxiv}, is today available dealing with all three cases in a comprehensive way.

The main contribution of this paper is a shadow fading channel model (LOS/OLOS model) based on real measurements in highway and urban scenarios distinguishing between LOS and OLOS. The model targets vehicular \emph{ad hoc} network (VANET) system simulations. We also provide a solution on how to incorporate the LOS/OLOS model in a VANET simulator. We model the temporal correlation of shadow fading as an auto-regressive process. Finally, simulation results are presented where the results obtained from the LOS/OLOS model are compared against the Cheng's model \cite{Cheng07}, which is also based on an outdoor channel sounding measurement campaign performed at $5.9$\,GHz. The reason to choose Cheng's model for comparison is that the Cheng's model do not classify measured data as LOS, OLOS and NLOS, but it represents both the small-scale fading, and the shadowing by the Nakagami-$m$ model.
   
The remainder of the paper is organized as follows. Section \ref{sec:Methodology} outlines the outdoor V2V measurements and explains the methods for separating LOS, OLOS and NLOS data samples which serves as first step to model the effects of shadow fading. It also includes the derivation of path loss and modeling of shadow fading in LOS and OLOS cases as log-normal distribution. The channel model is provided in section \ref{sec:Channelmodel}. First, the extension in traffic mobility models is suggested to include the effect of large-scale fading, and then the path loss model is presented and parameterized based on the measurements. VANET simulation results are discussed in Section \ref{sec:Networksimulations}. Finally, section \ref{sec:Conclusions} concludes the paper.

\section{Methodology}
\label{sec:Methodology}
\subsection{Measurement Setup}
Channel measurement data was collected using the RUSK-LUND channel sounder, which performs multiple-input multiple-output (MIMO) measurements based on the switched array principle. The measurement bandwidth was $200$\,MHz centered around a carrier frequency of $5.6$\,GHz and a total $N_f=641$ frequency points. For the analysis the complex time-varying channel transfer function $H(f,t)$ was measured for two different time durations: short term (ST), $25$\,s, and long term (LT), $460$\,s. The short-term and long-term channel transfer functions were composed of total $N_t = 49152$ and $N_t = 4915$ time samples, sampled with a time spacing of $\Delta t= 0.51$\,ms and $\Delta t= 94.6$\,ms, respectively. The test signal length was set to $3.2\,\mu$s. 

Two standard $1.47$\,m high station wagons, Volvo V70 cars, were used during the measurement campaign. An omni-directional antenna was placed on the roof of the TX and RX vehicles, taped on a styrofoam block that in turn was taped sideways to the shark fin on the center of the roof, and $360$\,mm from the back edge of the roof.  Videos were taken through the windscreen of each TX/RX car and GPS data was also logged during each measurement. Video recordings and GPS data together with the measurement data were used in the post processing to identify the LOS, OLOS and NLOS conditions, important scatterers, and to keep track of the distance between the two cars. The videos were synchronized to the measurements. 

\subsection{Measurement routes}

Eight routes in two different propagation environments were chosen with differences in their traffic densities, roadside environments, number of scatterers, pedestrians and houses along the road side. All measurements were conducted in and between the cities Lund and Malm\"{o}, in southern Sweden. 

\emph{Highway}: Measurements were performed when both the TX and RX cars were moving in a convoy at a speed of $22-25$\,m/s ($80-90$\,km/h), on a $2$ lane (in each direction) highway, between the cities of Lund and Malm\"{o}, Sweden. The traffic density was varying on both sides of the road from low to high traffic. Along the roadside there were trees, vegetation, road signs, street lights and few buildings situated at random distances. The direction of travel was separated by a ($\approx 0.5$\,m tall) concrete wall whereas the outer boundary of the road was guarded by a metallic rail. 

\emph{Urban}: Measurements were performed when both the TX and RX cars were moving in a convoy as well as in the opposite directions, in densely populated areas in Lund and Malm\"{o}. The TX and RX cars were moving with different speeds, between $0-14$\,m/s ($0-50$\,km/h), depending on the traffic situation. The streets were either single or double lane ($12-20$\,m wide) lined with $2-4$\,storied buildings. There were trees on either side and sidewalks on both sides of the streets. Moreover, there were road signs, street lights, bicycles and many parked cars, usually on both sides of the street. The streets were occupied with a number of moving vehicles as well as few pedestrians walking on the sidewalks.  

In total $3$ ST and $2$ LT measurements for highway, and, $7$ ST and $4$ LT measurements for urban-convoy were performed. During each measurement, the LOS was often obstructed by other cars, taller vans, trucks, buses or houses at the street corners. 

\subsection{LOS, OLOS and NLOS separation}
The measured channel transfer functions were first separated as LOS, OLOS and NLOS for all the measurements. The geometric information available form the video recording from the measurements has served as the foundation for the separation process. To distinguish the LOS samples from OLOS and NLOS samples, we define LOS condition as when it is possible for one of the cameras to see the middle of the roof of the other vehicle. Otherwise, the LOS is categorized as blocked. The blocked LOS situation is further categorized into two groups as defined above; OLOS and NLOS. The recorded videos contain 25 frames per second and are synchronized to the measurements. Therefore, a frame-by-frame evaluation of video data is performed to separate the data for LOS, OLOS and NLOS cases.  


\begin{table}[b!]
  \centering
  \caption{Distance traveled in LOS, OLOS and NLOS conditions.}
    \begin{tabular}{lllllll}
    \toprule
										& Scenario & Total & Min & Max & Mean & Median\\     
    \midrule
    LOS (m) 				& Highway & 6622 & 24.4 & 2157 & 299 & 125\\
										& Urban & 5477 & 0.95 & 519 & 84.6 & 35.3\\
    \hline
    OLOS (m) 				& Highway & 10752 & 18.6 & 2298 & 467 & 150\\
										& Urban & 2429 & 2.4 & 656 & 39.8 & 20.5\\
    \hline
    NLOS (m) 				& Urban & 415 & - & - & - & -\\
    \bottomrule
    \end{tabular}%
  \label{tab:LOS_NLOSV_Interval}%
\end{table}%

\begin{figure}
     \begin{center}
     \label{fig:Intervals}
     \includegraphics[width=0.5\textwidth]{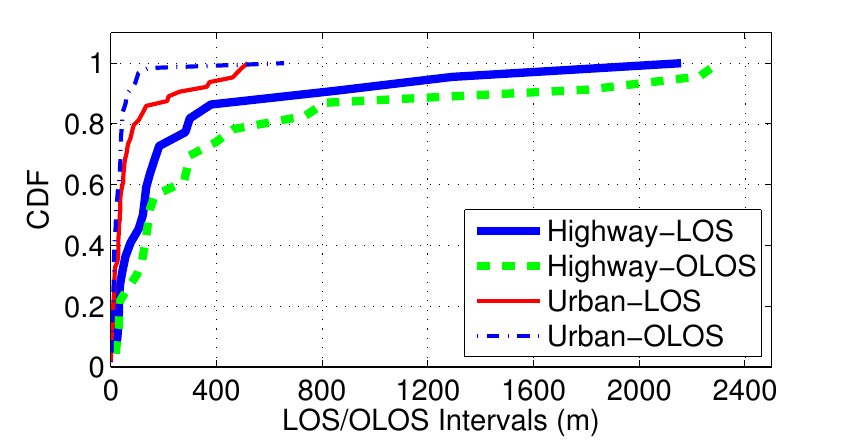}
    \end{center}
    \caption{%
        Cumulative Distribution Function (CDF) of LOS and OLOS distance intervals for all measurements; (a) highway scenario, (b) urban scenarios.
     }%
   \label{fig:Intervals_Highway_Urban}
\end{figure}

During the whole measurement run, the TX-RX link transited between LOS, OLOS and NLOS states a number of times, i.e., LOS-to-OLOS: $61$ times in urban and $23$ times in highway scenario, similarly, LOS-to-NLOS: $4$ times in urban and $0$ times in highway scenario. Each time the TX-RX pair is in either LOS, OLOS or NLOS state, it remains in that state for some time interval and travel a certain distance relative to the speed vehicles. In Table \ref{tab:LOS_NLOSV_Interval} the traveled distances for both the scenarios; urban and highway, are tabulated together with the distances where TX/RX were in LOS, OLOS and NLOS, respectively. The Cumulative Distribution Function (CDF) of these LOS/OLOS distance intervals are shown in Fig.~\ref{fig:Intervals_Highway_Urban}. No transition took place from OLOS-to-NLOS. The NLOS does not usually occur on highways, and the data samples for NLOS data in urban measurements are too few to be plotted as a CDF.

\begin{figure*}
     \begin{center}
        \subfigure[]{%
            \label{fig:mixModel_Highway}
            \includegraphics[width=.31\textwidth]{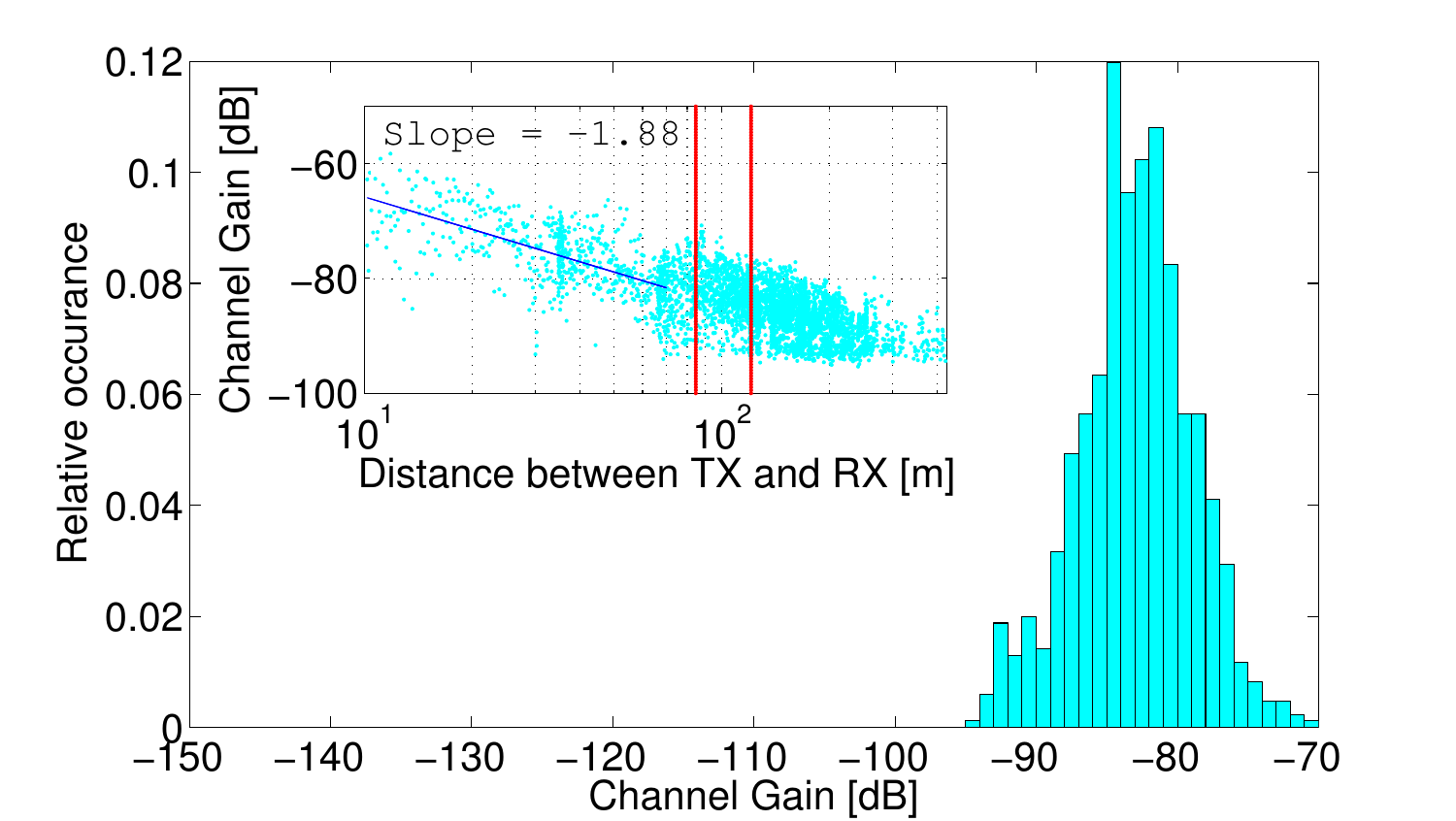}
        }%
        \subfigure[]{%
            \label{fig:pdfLOSvsNLOS_Highway}
        		\includegraphics[width=.31\textwidth]{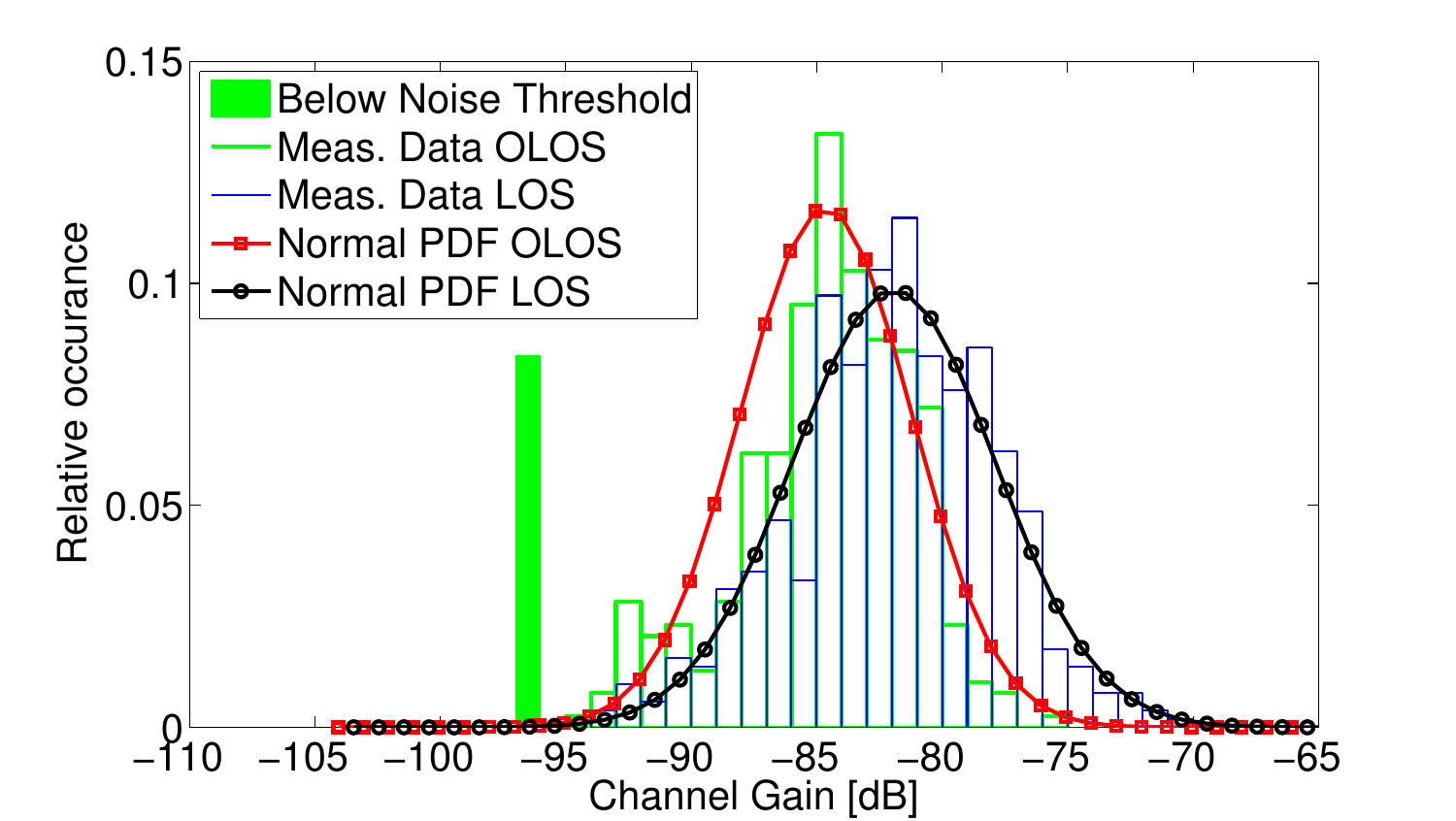}
  			}
  			\subfigure[]{%
  					\label{fig:cdfLOSvsNLOS} 
    				\includegraphics[width=.31\textwidth]{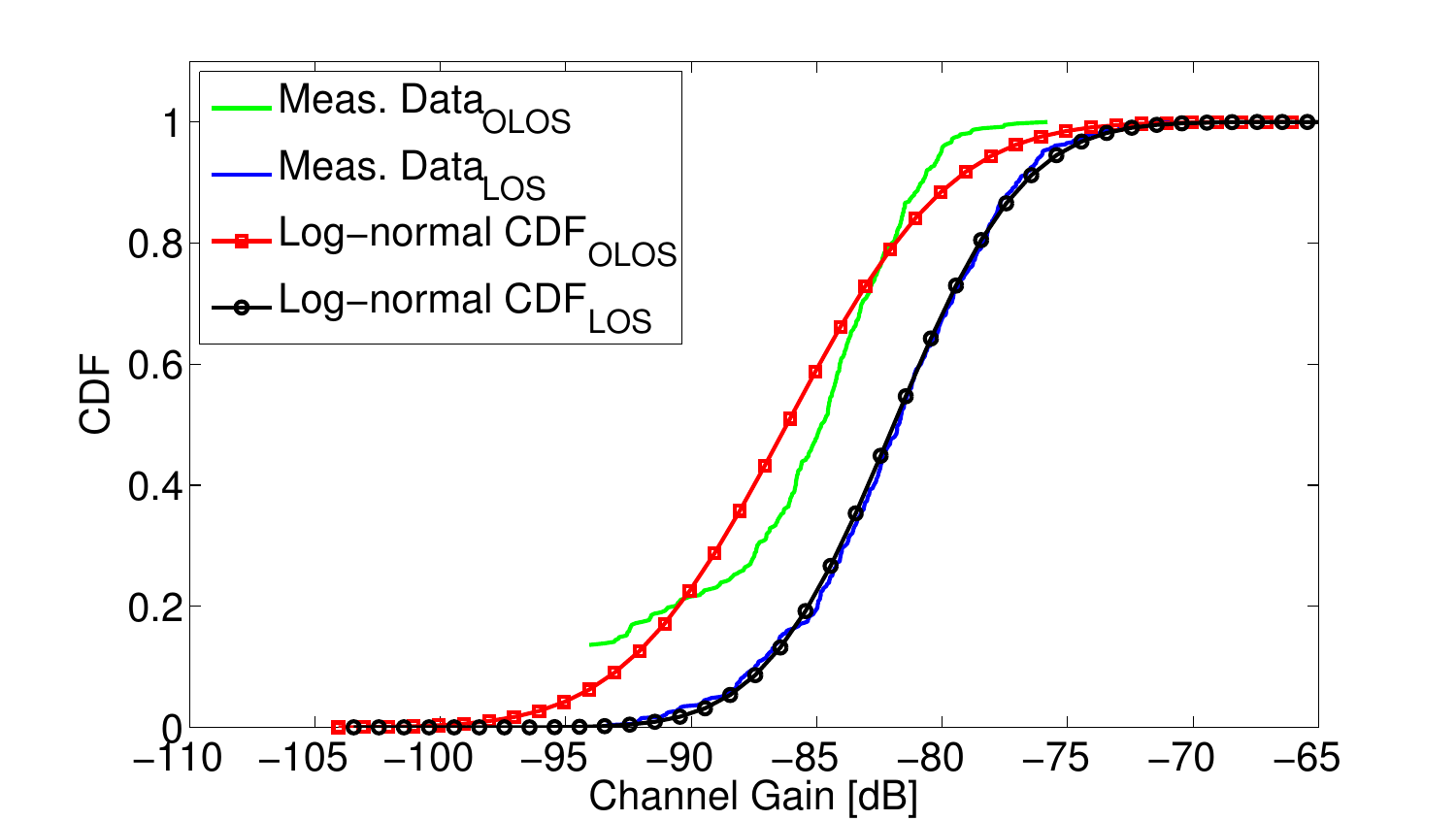}
  			}
        \\ 
    
        \subfigure[]{%
            \label{fig:mixModel_Urban}
            \includegraphics[width=.31\textwidth]{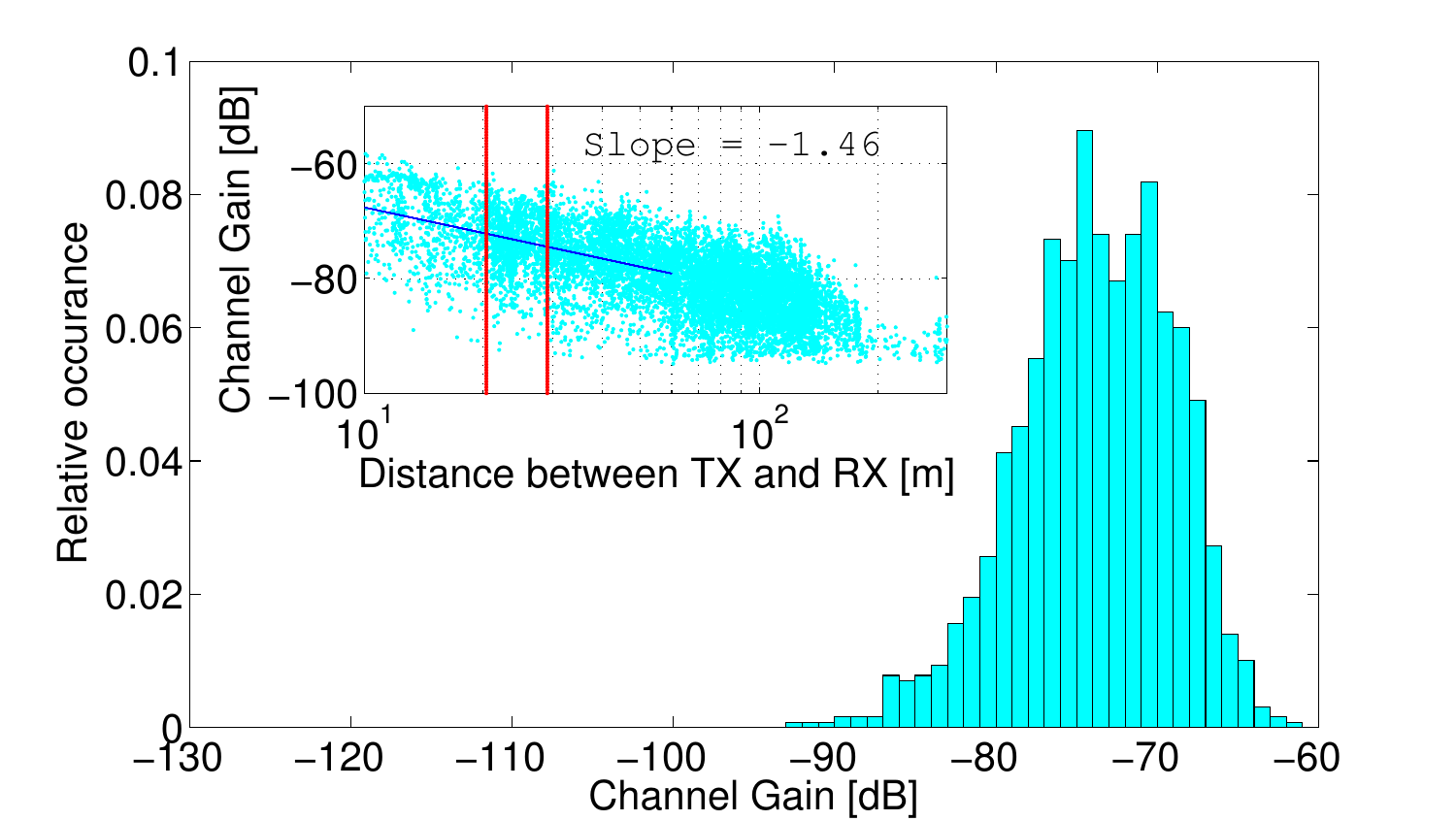}
        }%
        \subfigure[]{%
            \label{fig:pdfLOSvsNLOS_Urban}
        		\includegraphics[width=.31\textwidth]{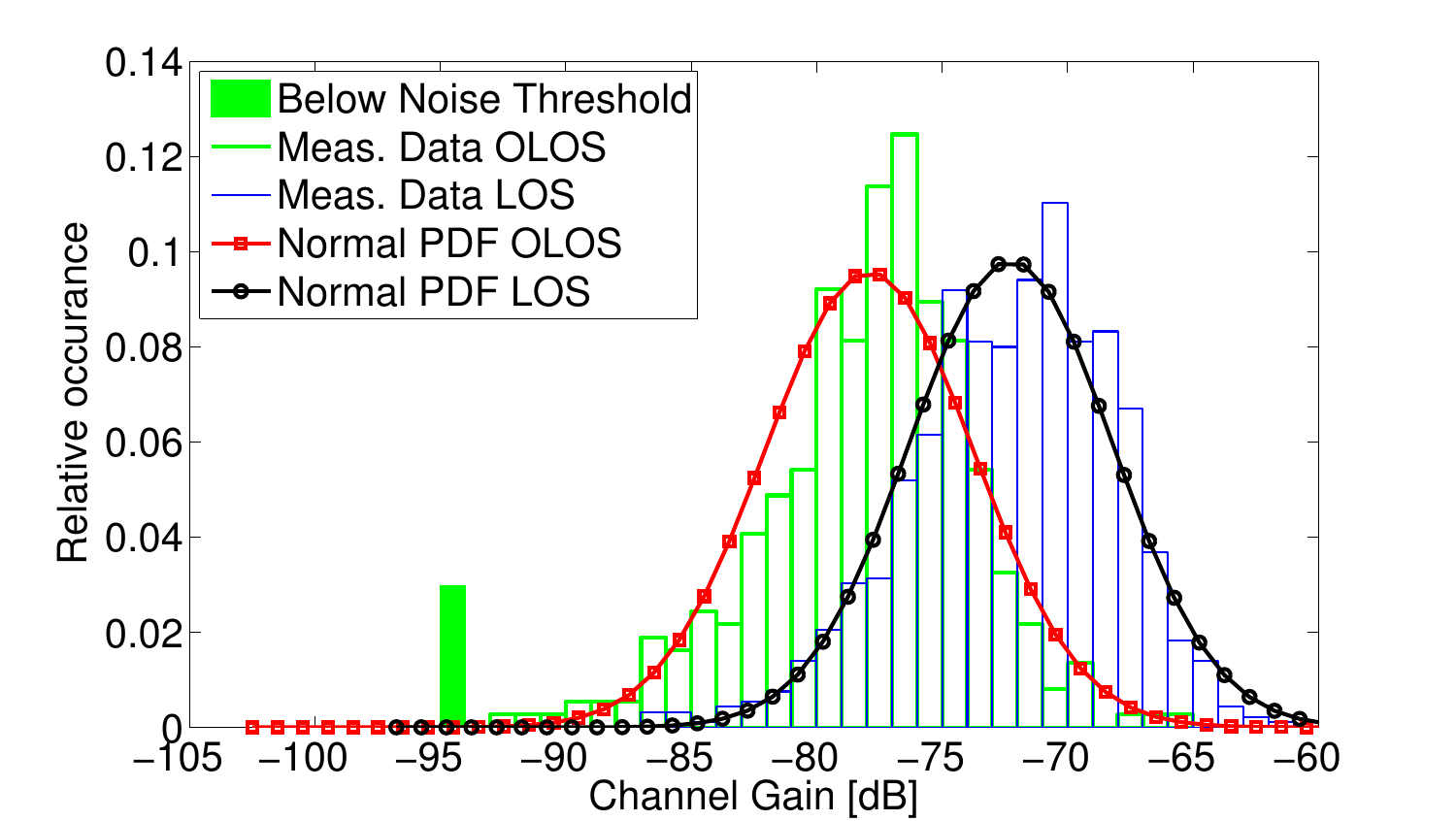}
  			}
  			\subfigure[]{%
  					\label{fig:cdfLOSvsNLOS_Urban} 
    				\includegraphics[width=.31\textwidth]{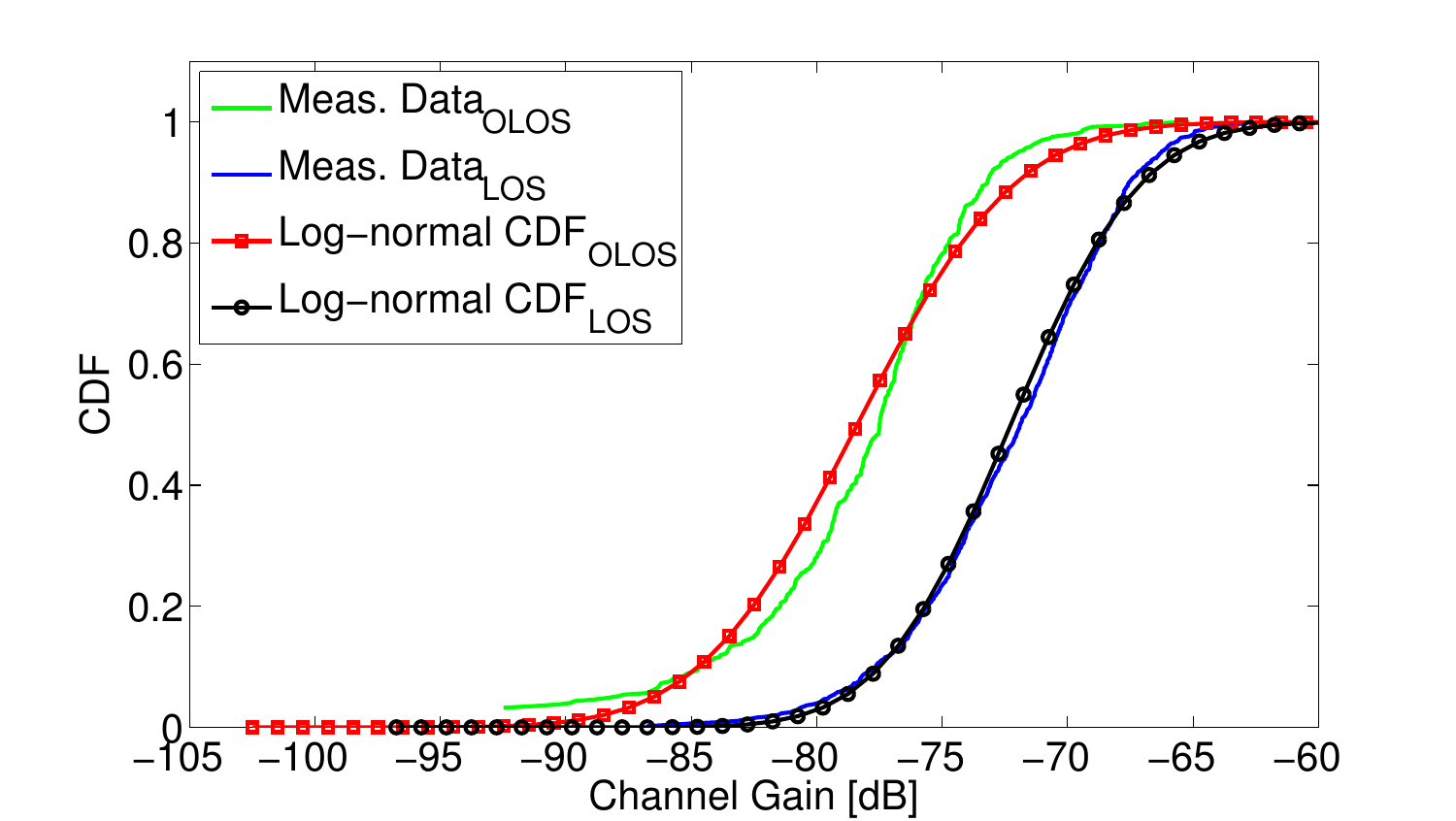}
  			}
    \end{center}
    \caption{Figures a, b, c represent highway data, and d, e, f represent urban data: In (a) and (d) inset plots shows channel gain for the overall measurement data as a function of direct distance between TX and RX. The slopes are provided up to the distance where no sample is below the noise. Large plot is the histogram of channel gains taken from log spaced distance bin, $20.4-29.1$\,m in (a) and $84.6-121$\,m in (d), marked by vertical lines in the inset plot. The LOS and OLOS data is treated together in these figure; (b) and (e) show histogram of the same channel gains shown in (a) and (d), when separated as LOS and NLOS, pdf fitting the Gaussian distribution; (c) and (f) show CDF fitting the log-normal distribution with $95$\% confidence interval.}%
   \label{fig:subfigures}
\end{figure*}

\subsection{Pathloss Derivation}

The time varying power-delay-profile (PDP) is derived for each time sample in order to determine the path loss. The effect of small scale fading is eliminated by averaging the time varying PDP over $N_{avg}$ number of time samples, the averaged-PDP (APDP) is given by \cite{karedal11} as
\begin{equation}
\label{eq:APDP}
P_h(t_k,\tau)=\frac{1}{N_{avg}}\sum_{n=0}^{N_{avg}-1}|h(t_k+n\Delta t,\tau)|^2,
\end{equation} 
for $t_k=0,N_{avg}\Delta t,...,\left\lfloor{{N_t}/{N_{avg}}-1}\right\rfloor N_{avg}\Delta t $, where $h(t_k+n\Delta t,\tau)$ is the complex time varying channel impulse response derived by an inverse Fourier transform of a channel transfer function $H(f,t)$ for a single-input single-output (SISO) antenna configuration. The $N_{avg}$ corresponds to the movement of the TX and RX by $s=15\lambda$ and is calculated by $N_{avg}=\frac{s}{v\Delta t}$, where $\Delta t$ is the time spacing between snapshots and $v$ is the velocity of TX and RX given in each scenario description. $N_{avg}\Delta t$ equals $32$\,ms and $71$\,ms for highway and urban scenarios, respectively, and they are chosen such that the wide-sense stationary (WSS) assumption is valid over $N_{avg}$ snapshots \cite{LauraPMIRC2012}.

The noise thresholding of each APDP is performed by suppressing all signals with power below the noise floor, i.e., noise power plus a $3$\,dB additional margin, to zero. The noise power is determined from the part of PDP, at larger delays, where no contribution from the transmitted signal is present. Thus, the zeroth order moment of the noise thresholded, small-scale averaged APDPs gives the averaged channel gain for each link as, 
\begin{equation}
G_h(t_k)= \sum_{\tau}P_h(t_k,\tau),
\label{eq:ChannelGain}
\end{equation}
where $\tau$ is the propagation delay.  


Finally, the antenna influence and other implementation losses such as cable attenuation and the effect of the low-noise-amplifier (LNA) were removed from the measured gains. The azimuth antenna pattern was almost omni-directional with variations of about $2$\,dB and peak gain of about $3.7$\,dBi, which was measured in an anechoic chamber. The distance dependent path-loss $PL(d)$ is then calculated using the following equation,

\begin{equation}
PL(d) = 2G_a-P_{IL}-10\log_{10}G_h(d),
\label{eq:Pathloss}
\end{equation}
where $G_h(d)$ is the distance dependent channel gain, which is obtained by matching the time dependent channel gain $G_h(t_k)$ to its corresponding distance $d$ between TX and RX at time instant $t_k$, $G_a$ is the antenna gain, and $P_{IL}$ is the implementation loss. 

GPS data, recorded during the measurements, was used to find the distance between TX and RX, which corresponds to the propagation distance of first arriving path for each time sample in the presence of LOS. The time resolution of the GPS data was limited to one GPS position/second. Thus, to make GPS data sampling rate equal to the time snapshots, interpolation of the GPS data was performed with a cubic spline method. The distance obtained from the GPS data was further validated, later, by tracking the first arrived MPC, in the presence of LOS, with a high resolution tracking algorithm \cite{karedal09}. 


\subsection{Large-scale or shadow fading}

As explained earlier the effect of small scale fading is removed by averaging the received signal power over a distance of $15$\,$\lambda$. The averaged envelope is a random variable due to the large-scale variations caused by the shadowing from large objects such as building, and vehicles. The most widely accepted approach is to model the large-scale variations with a log-normal distribution function \cite{book:Stuber,molisch05}. 

For the analysis of large-scale variations the distance dependent channel gain $G_h(d)$ is divided into log-spaced distance bins and the distribution of the data associated to each bin is studied independently with and without the separation of LOS, OLOS, and NLOS data. Before the data separation it was observed that a log-normal distribution does not provide a good approximation of the data, as often anticipated, as shown in Fig.~\ref{fig:mixModel_Highway}~and~\ref{fig:mixModel_Urban}. Moreover, an additional attenuation was observed, which made the spread in the channel gain large and the spread was different for different distance bins. The conclusion was drawn that the attenuation could possibly be associated to the LOS obstruction. Therefore, it was important to separate the data for the LOS, OLOS and NLOS situations. The analysis of the separated data sets showed that the large-scale variations for both LOS and OLOS can be modeled as log-normal distribution (see, e.g., Fig.~\ref{fig:pdfLOSvsNLOS_Highway}, \ref{fig:cdfLOSvsNLOS_Urban}) with an offset of almost $10$\,dB in their mean. In highway and open scenarios even higher losses are expected due to obstructed LOS especially when the two cars have communication distance less than $80$\,m. This observation is in line with the independent observations presented in \cite{Meireles10}. In which it is found that a single obstruction at the communication distance of about $10$\,m can cause an additional attenuation of about $20$\,dB.

%

The channel gain in the OLOS condition momentarily falls below the noise floor of the channel sounder and power levels of samples below the noise threshold can not be detected correctly. However, the OLOS data in each bin for shorter distances, with no missing samples, fits well to a log-normal distribution, and the assumption is made that the data continues to follow a log-normal distribution for larger distance bins where the observed data is incomplete. Moreover, the exact count of missing samples is also available, which can be used to estimate the overall data distribution. To get the mean and variance of Gaussian distributed LOS and OLOS data, the maximum likelihood estimates (MLE) of scale and position parameters from incomplete data were computed by using the method in \cite{Dempster77maximumlikelihood}, in which Dempster \emph{et~al.} presented a broadly applicable algorithm that iteratively computes MLE from incomplete data via expectation maximization (EM). 

\section{Channel Model}
\label{sec:Channelmodel}

In this section a shadow fading model (LOS/OLOS model) for VANET simulations is provided. This model targets network simulation, where there is a need for a realistic model taking shadowing effects into account but still with a reasonable complexity.

\subsection{Pathloss Model}

The measurement data is split into three data sets; LOS, OLOS and NLOS. The parameters of the path loss model are extracted only for the LOS and the OLOS data sets, whereas, not enough data is available to model the path loss for the third data set, NLOS. 

\begin{figure} 
\centering
  \includegraphics[width=.45\textwidth]{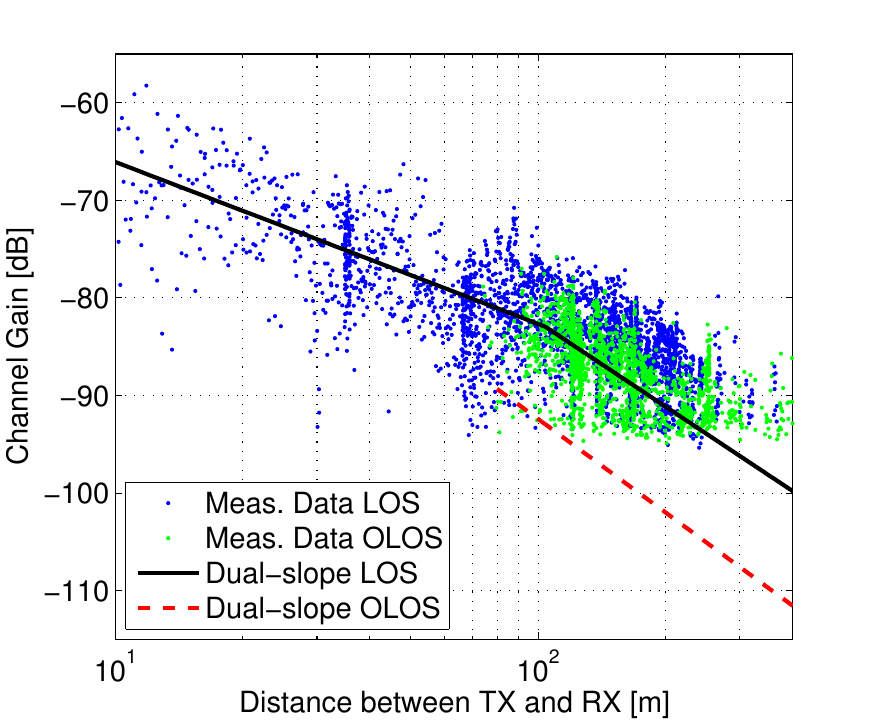}
  \caption{Measured channel gain for the highway environment and the least square best fit to the deterministic part of (\ref{eq:PowerLaw}).}
  \label{fig:PL_HiConvoy} 
\end{figure}

\begin{figure}
\centering
  \includegraphics[width=.45\textwidth]{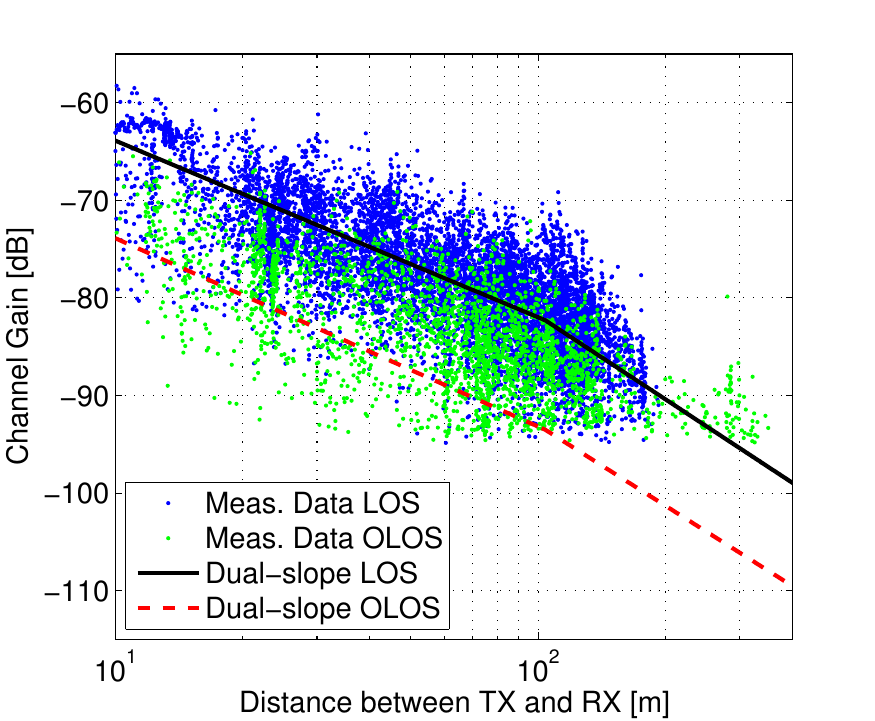}
  \caption{Measured channel gain for the urban environment and the least square best fit to the deterministic part of (\ref{eq:PowerLaw}).}
  \label{fig:PL_UrConvoy} 
\end{figure}

The measured channel gain for LOS and OLOS data for the highway and the urban scenario is shown as a function of distance in Fig.~\ref{fig:PL_HiConvoy} and \ref{fig:PL_UrConvoy}, respectively. A simple log-distance power law \cite{molisch05} is often used to model the path loss to predict the reliable communication range between the transmitter and the receiver. The generic form of this log-distance power law path loss model is given by,

\begin{equation}
PL(d)= PL_0+10n\log_{10}\left(\frac{d}{d_0}\right)+X_\sigma,
\label{eq:PowerLaw}
\end{equation}
where $d$ is the distance between TX and RX, $n$ is the path loss exponent estimated by linear regression and $X_\sigma$ is zero-mean Gaussian distributed random variable with standard deviation $\sigma$ and with some time correlation. The $PL_0$ is the path loss at a reference distance $d_0$ in dB.

In practice it is observed that a dual-slope model, as stated in \cite{Cheng07}, can represent measurement data more accurately. We thus characterize a dual-slope model as a piecewise-linear model with the assumption that the power decays with path loss exponent $n_1$ and standard deviation $\sigma$ until the breakpoint distance $(d_b)$ and from there it decays with path loss exponent $n_2$ and standard deviation $\sigma$. The dual-slope model is given by,

\begin{equation}
    PL(d)= 
\begin{cases}
    PL_0+10n_1\log_{10}\left(\frac{d}{d_0}\right)+X_{\sigma},& \text{if } d_0\le d\le d_b\\
    PL_0+10n_1\log_{10}\left(\frac{d_b}{d_0}\right)+ & \text{if } d > d_b\\
    		10n_2\log_{10}\left(\frac{d}{d_b}\right)+X_{\sigma}.
\end{cases}
\label{eq:dual-slope-PowerLaw}
\end{equation}

\begin{table}[htbp]
  \centering
  \caption{Parameters for the dual-slope path loss model.}
    \begin{tabular}{llllll}
    \toprule
								& Scenario & $n_1$ & $n_2$ & $PL_0$ & $\sigma$\\
    \midrule
    LOS 				& Highway &  -1.66 &  -2.88 &  -66.1 & 3.95 \\
								& Urban &  -1.81 &  -2.85 &  -63.9 & 4.15 \\
    \midrule
    OLOS 				& Highway &  - &  -3.18 &  -76.1 & 6.12 \\
								& Urban &  -1.93 &  -2.74 &  -72.3 & 6.67 \\
    \bottomrule
    \end{tabular}%
  \label{tab:pathloss_model_dual_slope}%
\end{table}%

The distance between TX and RX is extracted from the GPS data, which can be unreliable when TX-RX are very close to each other. Moreover, there are only a few samples available for $d<10$\,m, thus the validity range of the model is set to $d>10$\,m and let $d_0=10$\,m. The typical flat earth model consider $d_b$ as the distance at which the first Fresnel zone touches the ground or the first ground reflection has traveled $d_b+\lambda/4$ to reach RX. For the measurement setup the height of the TX/RX antennas was $h_{TX}=h_{RX}=1.47$\,m, thus, $d_b$ can be calculated as, $d_b=\frac{4h_{TX}h_{RX}-\lambda^2/4}{\lambda}=161$\,m for $\lambda=0.0536$\,m at $5.6$\,GHz carrier frequency. A $d_b$ of $104$\,m was selected to match the values with the path loss model presented in \cite{Cheng07}, implying a somewhat better fit to the measurement data.  

The path loss exponents before and after $d_b$ in (\ref{eq:dual-slope-PowerLaw}) are adjusted to fit the median values of the LOS and OLOS data sets in least square sense and are shown in Fig.~\ref{fig:PL_HiConvoy} and \ref{fig:PL_UrConvoy}. The extracted parameters are listed in Table~\ref{tab:pathloss_model_dual_slope}. For the highway measurements, OLOS occurred only when the TX/RX vehicles were widely separated, i.e., when $d>80$\,m, which means that there are too few samples to model the path loss exponent in OLOS for shorter distances. Thus, the path loss exponent for OLOS for shorter distances is not modelled. Whereas in practice, this is not always the case, the OLOS can occur at shorter distances if there is traffic congestion on a highway with multiple lanes. 

It is interesting to notice that there is an offset $\Delta PL_0$ in the channel gain for the LOS and OLOS data sets, which is of the order of $8.6-10$\,dB, and is very similar to the results that have previously been reported. In \cite{Boban11} an additional attenuation of $9.6$\,dB is attributed to the impact of vehicle as an obstacle. Meireles \emph{et~al.} in \cite{Meireles10} stated that the OLOS can cause $10-20$\,dB of attenuation depending upon traffic conditions, as the congested traffic cause large attenuation. Moreover, it is important to mention that the path-loss exponents less than $2$ have also been found in many other studies \cite{Herman2013,karedal11,molisch09-CommMag,Kunisch2008,Cheng07}. It is mainly due to the effect of two-ray reflection model in open and highway scenarios or due to wave-guiding in urban scenarios.

It is highly important to model the pathloss in the NLOS situation because power level drops quickly when the LOS is blocked by buildings. As mentioned above, the available measured data in NLOS is not sufficient to model the pathloss, therefore it is derived from available models specifically targeting similar scenarios, such as, \cite{Giordano10,Sallabi00,Mangel011-3} and COST 231-Welfish-Ikegami model (Appendix 7.B in \cite{molisch05}). Among these, Mangel \emph{et~al.} in \cite{Mangel011-3} presented a realistic and a well validated NLOS pathloss model which is of low complexity, thus, enabling large-scale packet level simulations in intersection scenarios. The basis for the pathloss equation in \cite{Mangel011-3} is a cellular model proposed in \cite{Sallabi00}, which is slightly modified to correspond well to V2V measurements. Validation of the NLOS model using independent V2V measurement data is done in \cite{Abbas2013ITST}, the results show that the model is a good candidate for modeling the pathloss in NLOS. For completeness Mangel's model \cite{Mangel011-3} used for the NLOS situation is given as follows,

\begin{equation*}
    PL(d_r,d_t,w_r,x_t,i_s)= 3.75+i_s{2.94}
\end{equation*}
\begin{equation}
    +
\begin{cases}
    10\log_{10}\left(\left(\frac{d_t^{0.957}}{({x_t}{w_r})^{0.81}}\frac{4\pi d_r}{\lambda}\right)^{n_{NLOS}}\right),& \text{if } d_r\le d_b\\
    10\log_{10}\left(\left(\frac{d_t^{0.957}}{({x_t}{w_r})^{0.81}}\frac{4\pi d_r^2}{\lambda d_b}\right)^{n_{NLOS}}\right), & \text{if } d_r > d_b\\
   
\end{cases}
\label{eq:NLOS-path-loss}
\end{equation}
where $d_r/d_t$ are distance of TX/RX to intersection center, $w_r$ is width of RX street, $x_t$ is distance of TX to the wall, and $i_s$ specifies suburban and urban with $i_s=1$ and $i_s=0$, respectively. In a network simulator, the road topology and TX/RX positions are known, thus, these parameters can be obtained easily. The pathloss exponent in NLOS is provided in the model as $n_{NLOS}=2.69$ and Gaussian distributed fading with $\sigma=4.1$\,dB. 

For larger distances ($d_r>d_b$) the model introduces an increased loss due to diffraction, around the street corners, being dominant. The NLOS model is developed for TX/RX in intersecting streets. If the TX/RX are not in intersecting streets but in parallel streets with buildings blocking the LOS then this NLOS model is not advisable. The direct communication in such setting might not be possible or not required but these vehicles can introduce interference for each other due to diffraction over roof tops. This propagation over the roof top can be approximated by diffraction by multiple screens as it is done in the COST 231 model. However, in \cite{Gaugel12} simulation results are shown which state that the pathloss in parallel street is always very high, $>120$\,dB. The value is similar to the one obtained with theoretical calculations for diffraction by multiple screens. As the losses for the vehicles in parallel streets are high, interference from such vehicles can simply be ignored.

\begin{figure}
     \begin{center}
        \subfigure[]{%
            \label{fig:Corr_HiConvoy}
            \includegraphics[width=0.52\textwidth]{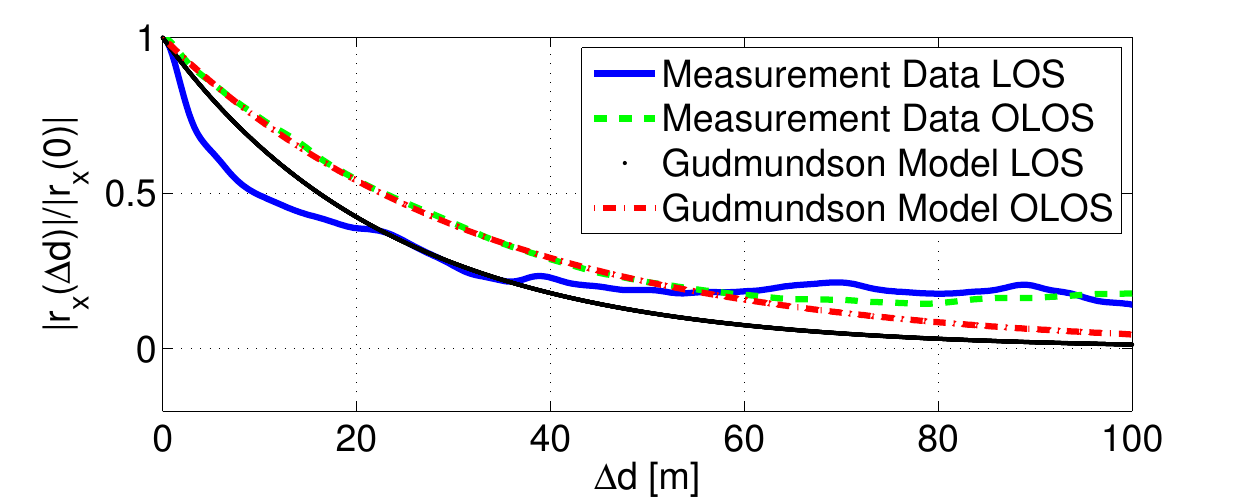}
        }%
        \\
        \subfigure[]{%
            \label{fig:Corr_UrConvoy}
            \includegraphics[width=0.52\textwidth]{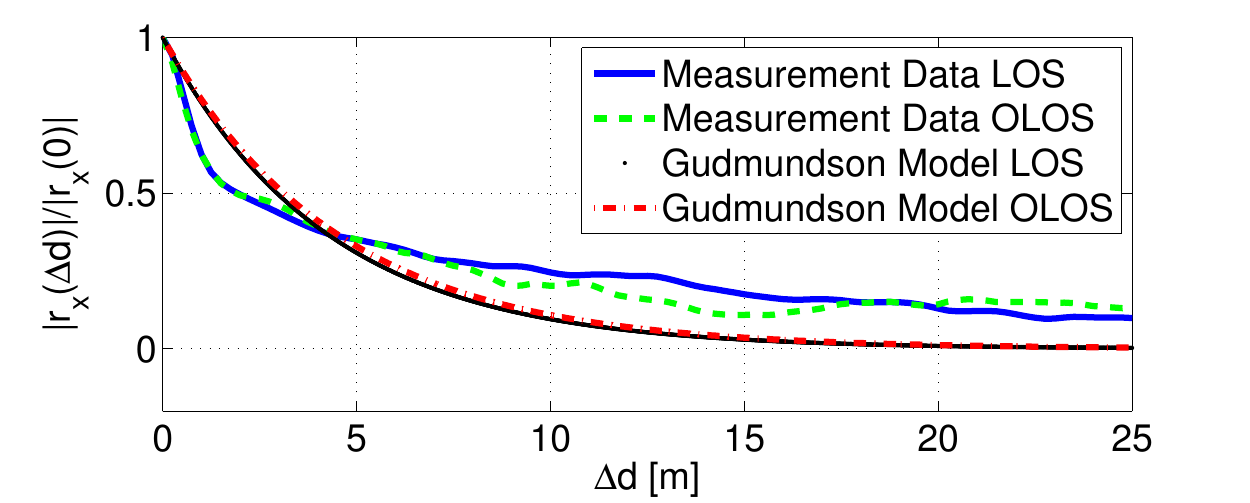}
        }%
    \end{center}
    \caption{%
        Measured auto-correlation function and model according to (\ref{eq:autocorr_model_Exp}) for LOS and OLOS data; (a) highway scenario, (b) urban scenarios.
     }%
   \label{fig:subfigures}
\end{figure}
\subsection{Spatial Correlation of Shadow Fading}
Once a vehicle goes into a shadow region, it remains shadowed for some time interval implying that the shadowing is a spatially correlated process. If a vehicle is in a shadow region there is possibility that its existence may not be noticed for some time. Hence, it is important to study the spatial correlation of the shadow fading as part of the analysis to find the average decorrelation distances.

The large-scale variation of shadow fading can be well described as a Gaussian random variable (discussed in section III). By subtracting the distance dependent mean from the overall channel gain, the shadow fading can be assumed to be stationary process. Then the spatial auto-correlation of the shadow fading can be written as,

\begin{equation}    
r_x(\Delta d)=E\{X_\sigma X_{\sigma}(d+\Delta d)\}.
\label{eq:autocorr}
\end{equation}

The auto-correlation of the Gaussian process can then be modeled by a well-known analytical model proposed by Gudmundson \cite{Gudmundson91}, which is a simple negative exponential function,

\begin{equation}    
r_x(\Delta d)=e^{{-|\Delta d|/d_c}},
\label{eq:autocorr_model_Exp}
\end{equation}

\begin{table}[htbp]
  \centering
  \caption{Decorrelation distances {$d_c$} for highway and urban scenarios.}
    \begin{tabular}{lllllll}
    \toprule
    \multicolumn{1}{l}{Scenario} & LOS & OLOS \\
    \midrule
    Highway & 23.3 & 32.5  \\
    Urban &4.25 & 4.5 \\
    \bottomrule
    \end{tabular}%
  \label{tab:pathloss_model_dual_slope}%
\end{table}%

where $\Delta d$ is an equally spaced distance vector and $d_c$ is a decorrelation distance being a scenario-dependent real valued constant. In the Gudmundson model, $d_c$ is defined as the value of $\Delta d$ at which the value of the auto-correlation function $r_x(\Delta d)$ is equal to $1/e$. The value of the decorrelation distance $d_c$ is determined from both the LOS and OLOS measured auto-correlation functions and are given in Table ~\ref{tab:pathloss_model_dual_slope}, for both the highway and urban scenarios, respectively. The estimated correlation distance is thus used to model the measured auto-correlation functions using (\ref{eq:autocorr_model_Exp}), and is shown in Fig.~\ref{fig:Corr_HiConvoy} and \ref{fig:Corr_UrConvoy}. 

Looking at decorrelation distance $d_c$, the implementation of shadow fading in a simulator can, if desired, be simplified by treating it as a block shadow fading, where $d_c$ can be assumed as a block length in which the signal power will remain, more or less, constant.  

\subsection{Extension in the Traffic Mobility Models}
The mobility models of today implemented in VANET simulators are very advanced, SUMO (Simulation of Urban Motility) \cite{SUMO2011} is one example of such an open source mobility model. These advanced models are capable of taking into account vehicle positions, exact speeds, inter-vehicle spacings, accelerations, overtaking attitudes, lane-change behaviors, etc. However, the ability to treat the vehicles as obstacles and model the intensity at which they obstruct the LOS for other vehicles is currently missing. Therefore, an extension for including shadowing effects in network simulators is provided herein, as stated in \cite{TaimoorHindai2013}. Since the vehicular mobility models implemented in the simulators give instantaneous information about each vehicle, the state of TX and RX vehicles can be identified by a simple geometric manipulation in the existing traffic mobility models as follows.
\begin{itemize}
\item Model each vehicle or building as a rectangle in the simulator.
\item Draw a straight line starting from the antenna position of each TX vehicle to the antenna position of each RX vehicle.
\item If the line does not touch any other rectangle, TX/RX has LOS.
\item If the line passes through another rectangle, the LOS is obstructed by a vehicle or by a building, the two cases are distinguished by using the geographical information available in the simulator.
\item Once the propagation condition is identified, the simulator can simply use the relevant model to calculate the power loss.
\end{itemize}
The impact of an obstacle is usually assessed qualitatively by the concept of the Fresnel ellipsoids. Only the visual sight does unfortunately not promise the availability of LOS, it is required that the Fresnel zone is free of obstacles in order to have the LOS \cite{molisch05}. The availability of LOS based on Fresnel ellipsoids depends very much on the information about the height of the obstacle, its distance from TX and RX, the distance between TX and RX as well as the wavelength $\lambda$. The information, if available in the traffic mobility simulator, should be utilized for the characterization of LOS and OLOS situation.

\section{Network simulations}
\label{sec:Networksimulations}

Finally, networks simulations are provided to show the difference between Cheng's Nakagami model \cite{Cheng07} and the channel model presented herein distinguishing between LOS and OLOS. The simulation scenario is a $10$\,km long highway with four lanes (two in each direction). The vehicles appear with a Poisson distribution with three different simulation settings for the mean inter-arrival time; $1$\,s, $2$\,s, and $3$\,s. The three different mean inter-arrival times yield three different vehicle densities, where $1$\,s corresponds is $\approx 100$\,vehicles/km, $2$\,s corresponds to $60$\,vehicles/km, and $3$\,s corresponds to $40$\,vehicles/km. Every vehicle broadcasts $400$\,byte long position messages at $10$\,Hz ($10$\,messages/sec) using a transfer rate of $6$\,Mbps and an output power of $20$\,dBm ($100$\,mW). The channel access procedure is carrier sense multiple access (CSMA), that has been selected as medium access control (MAC) for VANETs supporting road traffic safety applications. The vehicle speeds are independently Gaussian distributed with a standard deviation of $1$\,m/s, with different means ($23$\,m/s and $30$\,m/s) depending on lane. The vehicles maintain the same speed as long as they are on the highway. More details about the simulator can be found in \cite{KatrinPhDThesis}.
The shadowing based channel model LOS/OLOS model presented herein has been compared against Cheng's Nakagami model \cite{Cheng07} in the network simulations, where the latter is not capable of distinguishing between LOS and OLOS. Cheng's Nakagami model is also based on an outdoor channel sounding campaign, performed at $5.9$\,GHz in which the small-scale fading and the shadowing are both represented by the  Nakagami-$m$ model \cite{Cheng07}. The fading intensities, represented by the m parameter of the Nakagami distribution, are different depending on the distance between TX and RX. The m values and the path loss exponents are taken from data set 1 in \cite{Cheng07} to compute the averaged received power for Cheng's Nakagami model. 
\begin{figure}
\centering
  \includegraphics[width=.45\textwidth]{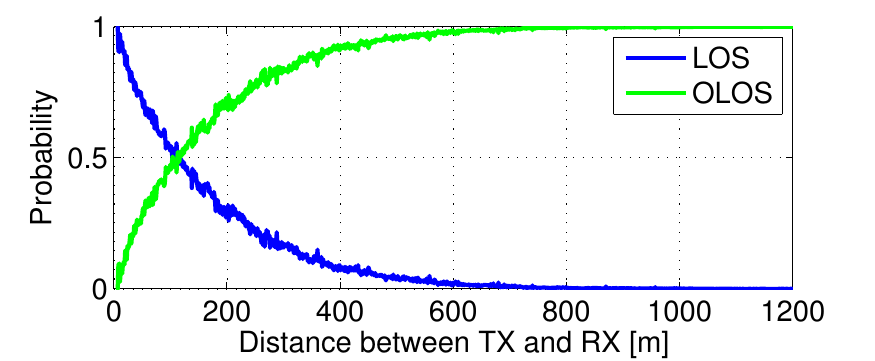}
  \caption{The probabilities of being in LOS and OLOS, respectively, depending on distance between TX and RX.}
  \label{fig:Prob-LOS-OLOS} 
\end{figure}

\begin{figure}
\centering
  \includegraphics[width=.45\textwidth]{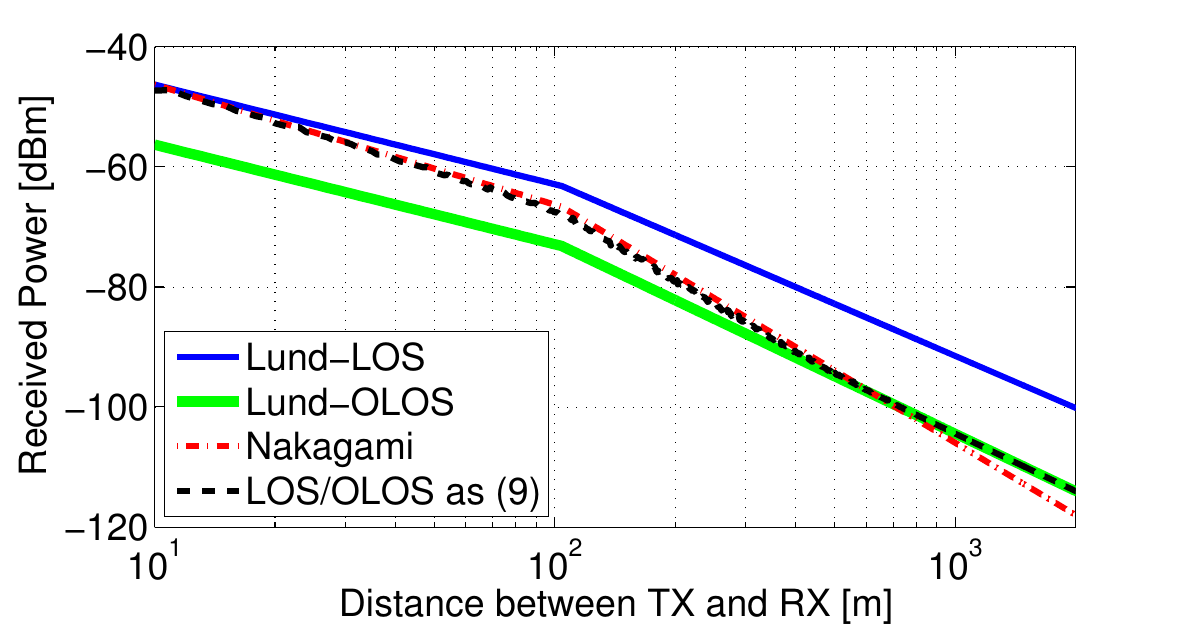}
  \caption{The averaged received power for the LOS/OLOS model and the Cheng's Nakagami model, respectively.}
  \label{fig:LundVsNakagami} 
\end{figure}
\begin{figure}
\centering
  \includegraphics[width=.45\textwidth]{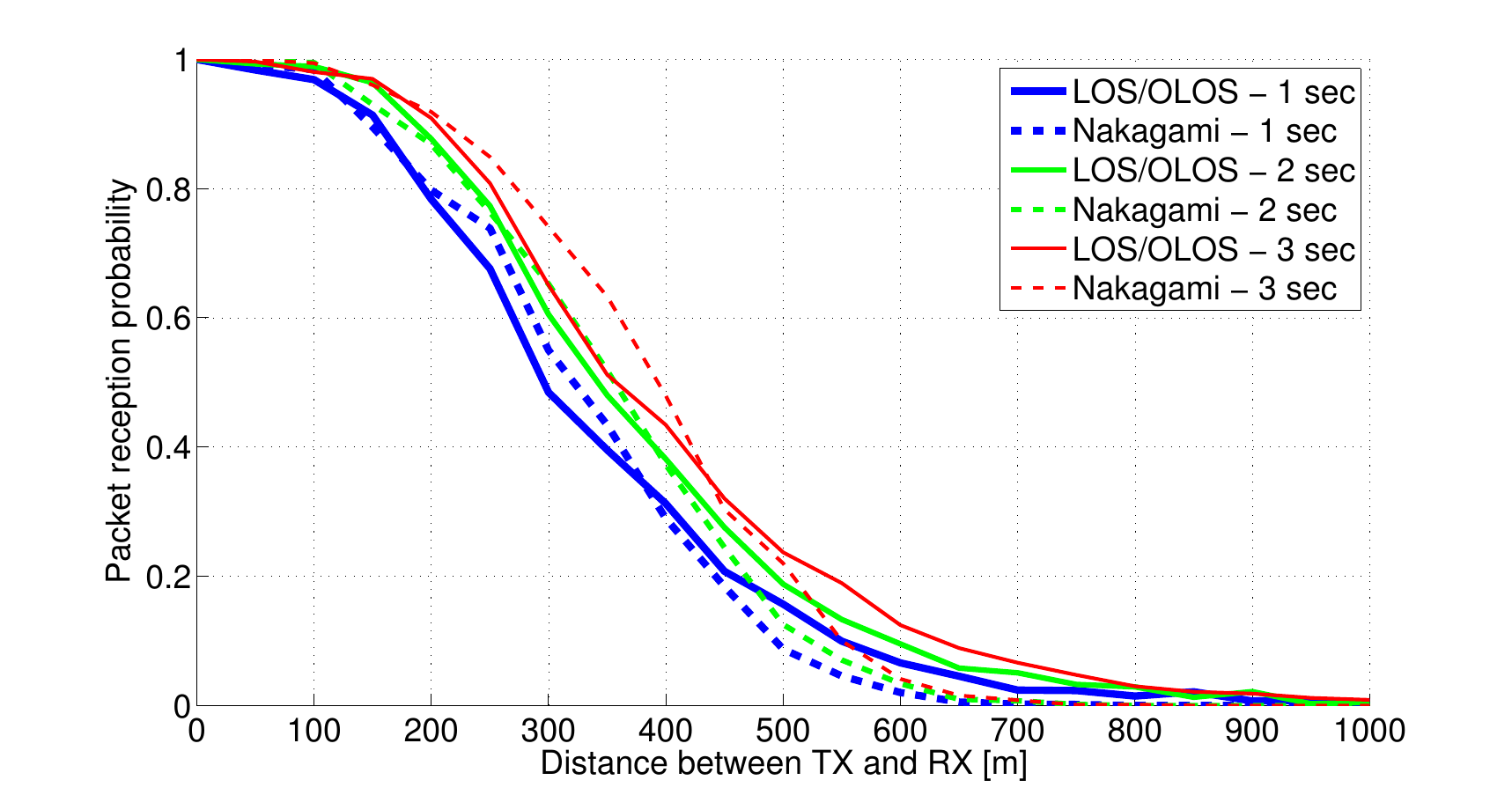}
  \caption{Packet reception probability for the two different channel models and for three different vehicle densities.}
  \label{fig:PKT_Rcep_Prob} 
\end{figure}

The averaged received power for the LOS/OLS and Cheng's Nakagami model is depicted in Fig.~\ref{fig:LundVsNakagami}. At shorter distances there is a little chance that another vehicle is between two communicating vehicles but as the distance increases the chances of being under OLOS either by vehicle, object, or due to the curvature of the earth, increases. The probabilities of being in LOS, $Prob(LOS|d)$, and being in OLOS, $Prob(OLOS|d)$, have been calculated from the network simulator for the highway scenario, as a function of distance and are depicted in Fig.~\ref{fig:Prob-LOS-OLOS}. To receive the averaged power as a function of distance similar to Cheng's model, these probabilities can be multiplied with the averaged received power for LOS, $P_{RX,LOS}(d)$, and OLOS, $P_{RX,OLOS}(d)$, using the following equation: 
\begin{eqnarray}\nonumber
P_{RX}(d)=Prob(LOS|d)P_{RX,LOS}(d)\\
+Prob(OLOS|d)P_{RX,OLOS}(d)
\label{eq:Model_LOS_OLOS_Prob}
\end{eqnarray}
By using (\ref{eq:Model_LOS_OLOS_Prob}) the averaged received power from the LOS/OLOS model coincides with Cheng's Nakagami model, see Fig.~\ref{fig:LundVsNakagami}, which is very interesting to notice. 
In Fig.~\ref{fig:PKT_Rcep_Prob}, the packet reception probability is depicted for the two channel models; LOS/OLOS model and Cheng's Nakagami model, respectively, and for three different vehicle densities. On the x-axis, the distance between TX and RX is shown. For detailed analysis six pairs of vehicles, three pairs in each direction are studied in the simulations.  Each pair travels in the same direction in different lanes with different speeds, where the vehicle with high speed will pass by the vehicle with the low speed. The selection of pairs where done to study individual performance of the vehicles. It should be noted that exactly the same number of communicating vehicles has been used for the different vehicle densities for every channel model, i.e., the same TX-RX pair are studied using both channel models. When TX and RX are close to each other, i.e., within $100$\,m, the two channel models perform equal. As the distance increases to $200-400$\,m, the vehicles exposed to the Nakagami model are experiencing a better packet reception probability. The vehicles exposed to the Nakagami model reach a packet reception probability of zero at around $700$\,m, whereas this is reached above $1000$\,m for the LOS/OLOS model. Real measurements \cite{AlexThesis} also show such a behaviour with occational successful packet transmission at larger distances. This also implies that the LOS/OLOS model contributes to interference at stations situated further away, which is in line with what is seen on the averaged received power in Fig.~\ref{fig:LundVsNakagami}. Here the OLOS part has a stronger signal than Cheng's Nakagami model above $700$\,m.  
\begin{figure}
\centering
  \includegraphics[width=.45\textwidth]{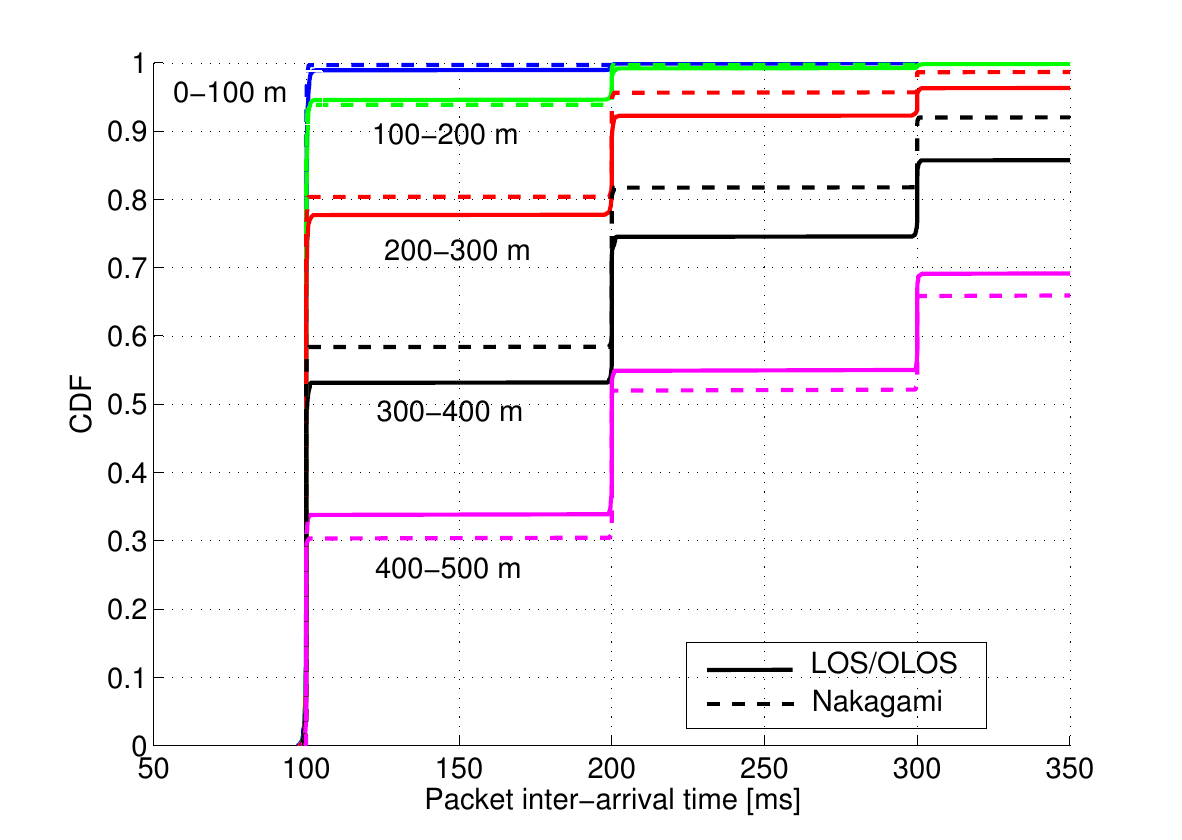}
  \caption{CDF for the packet inter-arrival time for a vehicle density of $40$ vehicles/km for the LOS/OLOS and Nakagami models, respectively.}
  \label{fig:PKT_time_40} 
\end{figure}

\begin{figure}
\centering
  \includegraphics[width=.45\textwidth]{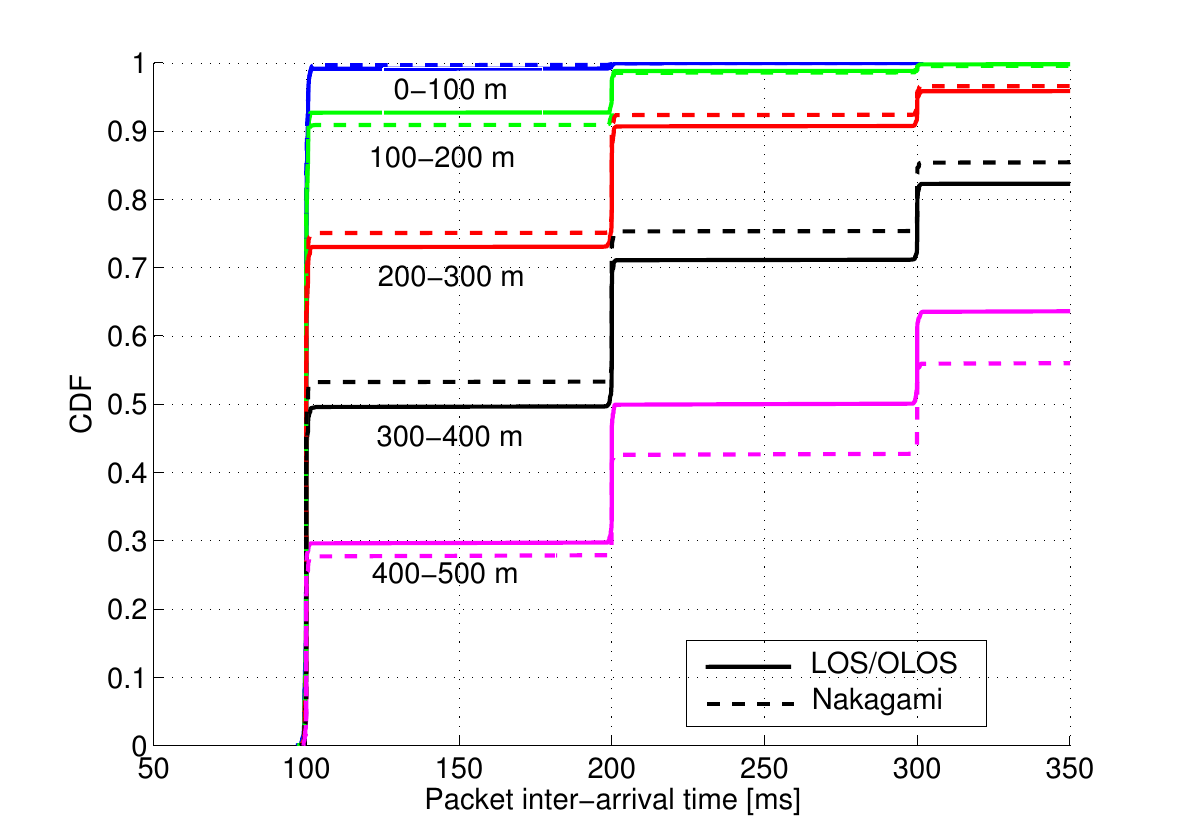}
  \caption{CDF for the packet inter-arrival time for a vehicle density of 60 vehicles/km for the LOS/OLOS and Nakagami models, respectively. }
  \label{fig:PKT_time_60} 
\end{figure}

\begin{figure}
\centering
  \includegraphics[width=.45\textwidth]{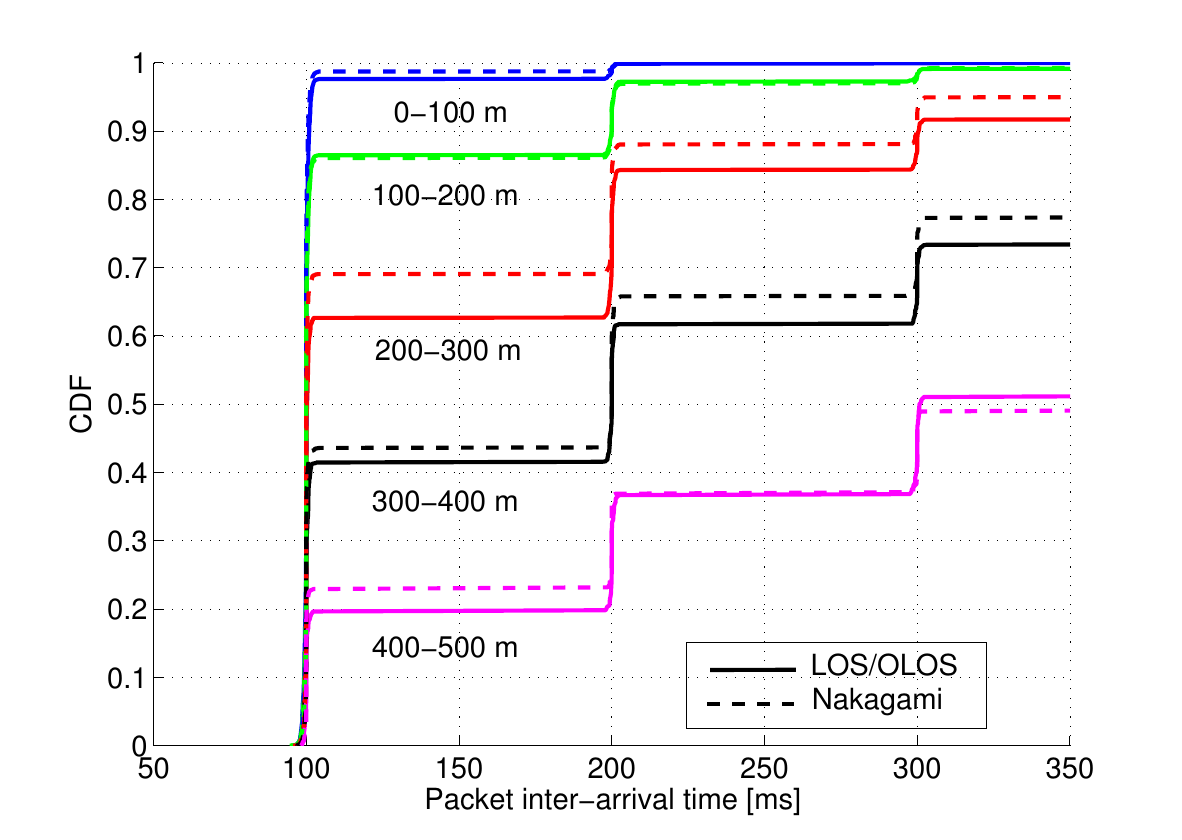}
  \caption{CDF for the packet inter-arrival time for a vehicle density of 100 vehicles/km for the LOS/OLOS and Nakagami models, respectively.}
  \label{fig:PKT_time_100} 
\end{figure}

In Fig.~\ref{fig:PKT_time_40}, the CDF for packet inter-arrival time for a vehicle density of $3$\,seconds ($40$\,vehicles/km) is depicted.  The received packet inter-arrival time is the time that has elapsed between two successfully received packets from a specific TX that RX is listening to. Around every $100$\,ms, the RX is expecting a new packet from a specific TX. The period is not exactly $100$\,ms due to channel access delays caused by, e.g., backoff procedures \cite{IEEE80211p}. The different lines in the figure represent different distance bins. When TX and RX are within $100$\,m of each other, the RX can expect a new packet every $100$\,ms and the channel models only differ slightly in performance. When distance increases to $200-400$\,m, the vehicles under the treatment of the Nakagami model are experiencing a better packet inter-arrival times, which was also reflected in Fig.~\ref{fig:PKT_Rcep_Prob}. In the distance bin $300-400$\,m, studying the vehicles under Nakagami, in almost $60\%$ of the cases there are no packets lost between two successful receptions and in about $25\%$ of the cases, a single packet is lost between successful receptions. As the distance increases to above $400$\,m, stations under the LOS/OLOS model are having slightly better packet reception probabilities. 
The CDF for packet inter-arrival times for a vehicle density of $2$\,s ($60$\,vehicles/km) is depicted in Fig.~\ref{fig:PKT_time_60}. Here, it is seen that the gap in performance between the two models has decreased since there are more transmissions on the channel and the overall interference has increased. 
In Fig. \ref {fig:PKT_time_100}, CDF for packet inter-arrival times for the highest vehicle density case is depicted. Still, the stations under the treatment of Nakagami are experiencing a better packet inter-arrival pattern, especially for distances between $200-300$\,m. 

\section{Summary and Conclusions}
\label{sec:Conclusions}
In this paper, a shadow fading model based on measurements performed in urban and highway scenarios is presented, where a separation between LOS, obstructed LOS by vehicle (OLOS) and obstructed LOS by building (NLOS), is performed. In the past, despite extensive research efforts to develop more realistic channel models for V2V communication, the impact of vehicles obstructing LOS has largely been ignored. We have observed that the LOS obstruction by vehicles (OLOS) induce an additional loss, of about $10$\,dB, in the received power. Network simulations have been conducted showing the difference between a conventional Nakagami based channel model (often used in VANET simulations) and the LOS/OLOS model presented herein. There is a difference in the performance of the two channel models. However, depending on the evaluated VANET application the obstruction of LOS cannot be ignored and there is a need for a LOS/OLOS model in VANET simulators. The LOS/OLOS model is easy to implement in VANET simulators due to the usage of a dual-piece wise path loss model and the shadowing effect is modeled as a log normal correlated variable with a mean determined by the propagation condition (LOS/OLOS).


\bibliographystyle{IEEEtran}
\nocite{*}
\bibliography{ShadowFading_Pathloss_model_Version_Hindawi.bbl}

\end{document}